# Mikhail Vasil'evich Lomonosov

## *Meditations on Solidity and Fluidity of Bodies* [1]

*for the solemn celebration of the namesake of Her Majesty, all the most beautiful and most powerful, the Grand Empress Elizaveta Petrovna, the autocrat of the All-Russia, read in the public meeting of the Imperial Academy of Sciences on September, 6th day of 1760, by the collegiate adviser and Professor and member of the Royal Swedish Academy of Sciences, Mikhail Lomonosov*

[ *Meditationes De Solido Et Fluido* ]

[ *Рассуждение о Твердости и Жидкости Тел* ]

§1

Everyone knows how much the solidity and liquidity of bodies depend on differences of heat and cold. Now when the partner in our studies, Professor Braun[2], has presented a report on his experiments on freezing of mercury in the past severe winter and has explained them, I consider it proper to present my own thoughts on the cause for cohesion in various bodies, so that one may obtain a clear and general idea of the freezing and thawing of sensible bodies[3] when they become solid or liquid, and by this public presentation of our works to express general and most humble reverence to this solemn festival.

§2

In studying the general reason for the cohesion of particles, I should first of all warn those who have not been interested in the subject and have been satisfied with speculative idea of just an attractive force, naively considering such as free of any striking contact. Therefore, I assert that I cannot acknowledge and accept such views because of my firm proffs, which, unless I am deceived, are new.

§3

If there really were an attractive force, it would have to be innate in bodies as a reason for the production of motion. But motion in a body is also produced by collision and bouncing, which is obvious to everyone. Thus, there are two direct and, moreover, opposite reasons for the motion: for what can be more contrary to attraction than repulsion? But directly opposite causes should lead to opposite actions. There are feeble examples which apparently may seem contradictory to this like, that animals die equally of strong heat and cold - but the reasons for that are essentially remote and indirect, and sometimes contradictory, while the immediate and direct cause of death is the suppression of flow and circulation of the blood and other vital fluids. (Newton himself stated just direct action and did not accept the attractive force in his lifetime, but after his death was made its unwilling defender by the overly great zeal of his epigones[4] ). And so, if the motion in bodies is due to attracting force, then it cannot be produced by collision or repulsion. But this is completely false, since repulsion truly produces motion and hence there is no true and undoubted attractive force in bodies.

---

[1] Translated to English from Russian version of Ref. [1] by V.Shiltsev; footnotes and commentary by V.Shiltsev
[2] Joseph-Adam Braun (1712?—1768), member of St.Petersburg Academy of Sciences since 1748; Lomonosov presented his "Meditations on Solidity and Fluidity of Bodies" in the public meeting of the Academy on September 6 (o.s.), 1760, after the speech in Latin read by J.-A. Braun, which was later published as "De admirando frigore artificiali, quo mercurius est congelatus" ("About the amazing cold, produced experimentally and which the mercury has frozen from", Petropoli, 1760) in Latin and in Russian [3].
[3] Here and below "sensible bodies" – "macroscopic bodies", contrary to "insensible particles" – microscopic, not detectable directly

§4

If we still concede that there is a true attractive force in bodies, then body $A$ attracts to itself another body $B$, that is, makes it move it without any contact, and for this there is no need for $A$ to touch $B$ and hence it is not necessary that $A$ moves toward $B$; but as there is no need for any motion of body $A$ in any direction to make body $B$ to move, it follows that $A$, even being completely immobile, moves $B$. Thus, $B$ obtains for itself something new, namely, the motion toward $A$ which it did not have before. But as the gist of all changes that occur in nature is that as much as is taken away from one body, so much is added to another; then, as much matter receded in one place, so much augmented in another place; as many extra hours one devotes to wakefulness, so many hours are to be taken away from sleep. This general law of nature extends in the laws of motion, for a body moving another by its force loses in itself as much as it passes to another which gets the motion. Hence, by this general law the motion of body $B$ to body $A$ is initiated by and taken away from body $A$. But as nothing can be taken away from nothing, it is necessary for body $A$ to be in motion when it attracts another body $B$ toward itself. But from preceding arguments it was concluded that body $A$ may be without any motion when it attracts another body $B$ to itself. And therefore, body $A$ can be in motion and stand perfectly still simultaneously. But as this contradicts itself and opposes the general philosophical statement that *the same thing can not be and not be at the same time*, then for the sake of truth, the original and direct attractive force does not occur in nature.

§5

Thus, it follows that the particles comprising the sensible bodies are hold in unity by impacts, or, better say, by pressure of some fluid matter surrounding them which is inactivated at their mutual contact. Therefore, it is necessary to consider how this fluid matter is squeezing the particles which compose the body to cohesion and then to explain briefly the nature of solid and liquid bodies.

---

[4] Reference to disciples and followers of I. Newton, in particular Roger Cotes; who was the editor of the 1713 edition of the " *Philosophiae naturalis principia mathematica"* and provided a long anti-Cartesian and anti-Leibnizian preface, where the Newton's hypothesis about the direct action of bodies at a distance was presented as the final and complete solution of the question and weight was declared as a general property of matter.

§6

Would not someone here ask me to name the reason, or what is the material and the nature of the coalescence of indivisible particles in the corpuscles compressed by the liquid matter surrounding them? Would not one say that I am now obliged to recognize the existence of attractive forces? By no means. Anyone who knows the difference between the essential invariable attributes of bodies and their variable qualities can see clearly that the reasons for these cannot be demonstrated nor should we ask what needs to be in things for their existence: for example, why a triangle has three sides, why a body has extension, and other similar questions. For the cause of cohesion should be sought where we see that insensible particles either coalesce or get free from bonding; or where the force of their cohesion either increases or decreases. Then one can only ask why this is so and not otherwise. As cohesion of the insensible particles which compose corpuscles is not considered variable, then the reasons for this also should not be asked. A philosophical *principle*, called a *satisfactory reason* does not extend to the essential properties of a body. From such incorrect usage comes extended discussion in scientific circles on simple substances, that is, of particles which do not have any extension. If extension is an essential attribute of a body, without which this body cannot exist, and almost all the virtue of the body 's definition consists in extension, then questions and quarrels concerning the non-extensive particles of an extensive body are futile; as the best method in such a case one should draw proofs from definitions instead of finding the proofs of the definition.

§7

Considering the particles' contact surfaces, I see at once a number of different shapes, ascribed by many physicists to insensible particles without success - though their ambition is praiseworthy, for the studies of the primary particles are as necessary as the existence the particles themselves. Since bodies cannot be composed other than from insensible particles, and without studying them the deepest physics understanding is possible. Seeing just the surface of a clock, how can one know by what forces they are moved and show the time separated into different equal parts. Without knowing the structure of the internal insensible particles, physicists and especially chemists would be in the dark. I do not count myself among the hopeless ones who are called discreet physicists and strive to know the shapes of insensible particles. Awry shapes of the particles: wedges, needles, hooks, rings, bubbles[5], and many others baselessly imagined and chosen without any basis by others do not deter me from studying particles which escape the view because of their smallness, unavailable to physical apparatus, and over twenty years of frequent discussions on the subject and correlating with experiments I have concluded that the nature is satisfied with sphericity only, the eases the task of those studying her secrets.

---

[5] Here Lomonosov derides ideas about the form of atoms widespread in the XVII century and in the beginning of the XVIII century, such as those of Pierre Gassendi (1592-1655) who believed that atoms possess very diverse forms - oval, oblong, pointed, angular, hooked, etc. - and that the form of atoms and their arrangements explain hardness and liquidity, adhesion and other properties of bodies. Similar ideas were held by Robert Boyle (1627-1691), Honore Fabry (1608-1688) and many others. Lomonosov expressed his Ironic attitude to this kind of ideas even in his early works – see, e.g. "*44 Notes on the Adhesion of Corpuscles*" (1743-1744, see in Ref. [2], vol.I, pp. 265-267).

§8

But having just probability, though very high, in fact, is not enough for me; it is also not enough that some intelligent people and most notable heroes of the learned world, consider the particles of matter to be spherical, especially those of liquids (and all solid bodies turn into liquids by the force of fire); I am not satisfied also with the reasoning by analogy that all natural bodies tend to be spheres and prefer this form - from the greatest even to the smallest, from the chief bodies of this world, as our earth, to the finest and simplest spheres invisible to the naked eye, which make up blood. I do not dwell on the fact that the parts of animals and plants, seeds and fruits have largely a circular shape, that the more finely divided all liquid materials are, the more spherical they become, I do not take into the consideration the evidences of the unnumbered multitude of spherical raindrops - I instead rest foundations of my proof on rigorous mathematics.

§9

I have shown before * [6] that Aristotelian elementary fire, or, according to the new scholarly style, special calorific matter, which is transferred from body to body and wandering, travels without the least probable cause and is only a fiction: and altogether established that fire and heat consist of a rotary motion of particles, especially of the corpuscles which make up bodies. That my system has been defended from unfounded objections, and attempted criticism has been presented in vain ** [7]. And above all, it has gotten new solid confirmations ***[8]. That's why, without hesitation, I use it as the basis for proof of the sphericity of the insensible particles which make up bodies.

\* In diss. De caus. Caloris et frig., Nov.C., t. I.
\*\* Vide Nouvelle Bibl., germ., t.VI.
\*\*\* In oratione de origine lucis, p.12.

§10

So, as the insensible particles of warm bodies perform rotary motion, let us assume that the particles of warm bodies are not spheres, but of some other shape, for example, cubic; from this would follow that while rotating, they are sometimes in contact by planes and at times by corners. From this it should follow: 1) that the cohesion of particles that makes bodies solid, is changing every moment, for on contact by corners little or nothing can hold one to the other; 2) all the diagonals and all the lines which form corners of the corpuscles are of different lengths, so that from moment to moment there should be a change of the size of sensible bodies and there would be a continuous vibration which would be stronger, the warner the body. But as evidenced by our senses neither of these effects is found in bodies, therefore, no cornered shapes or any other which has unequal diameters are possible in warm bodies, i.e. all bodies; that is, all shapes are impossible except the spherical.

---

[6] "Meditationes de caloris et frigoris causa (Meditations on the cause of heat and cold)" published in the *New Commentaries of the Petersburg Academy (Novi Commentarii Academiae Scientiarum Imperialis Petropolitanae*, t. I, 1750, pp. 206-229); see in Ref. [2], vol.II, pp. 7-55.

[7] Reference to Lomonosov's article addressing the objections of German critics of his molecular-kinetic theory of heat - "The Discourse on the Responsibilities of Journalists when Presenting Their Works for the Maintenance of Freedom of Philosophy," published in magazine *Nouvelle Bibliotheque germanique, ou d'histoire litteraire de l'Allemagne, de la Suisse et des pays du Nord* , vol. VII, part 2, pp.343-366 (1755, in French); see in Ref.[2], vol.III, pp. 201-232.

[8] Lomonosov's views on the structure of matter and the nature of the ether are set forth in the "Oratory of the Origin of Light, Presenting a New Theory of the Colors" (1756, see in Ref.[2], vol.III, pp. 316-344).

§11

All sensible bodies of any shape, put on a balance in equilibrium with weights in all positions, remain without fail in equilibrium; for example, marble or metallic pyramids placed on their bottom or on their ends, on their sides or on the angles, never show either increase or decrease in heaviness. These experiments, though very simple and known to everyone, yet are very important in the present case. We neglect and slide past many such simple and everyday effects which in the study of nature point the way to great discoveries, and we undertake difficult experiments, forgetting the very famous statement that is simple and based on irrefutable mathematics, that every substance equals its own size, on which almost all mathematics is founded. From the above mentioned every day and very simple art, it follows that all particles composing a body are in essence spherical shapes. For the action of gravitational material always and in the same way affects the impenetrable surfaces of the corpuscles in any position of these bodies and if shapes of the particles were not spherical, then all positions of the surface should differ for the gravitational material, with different forces and different weights. Thus, the particles composing bodies through which gravitational material cannot pass and only strikes the surface should be spherical.

§12

Having shown the spherical nature of the particles composing sensible bodies, where can we find contact surfaces? For spheres do not touch one another except at one point. To resolve this question satisfactorily it is necessary for me to determine the surface of contact (which should better be called the surface of cohesion), namely, that it is a circle whose diameter is the line *BID* [Fig. 1] between the particles *A* and *C* which are in contact, whose periphery is occupied by the very small spheres *B*, *D* of the pressing fluid matter, not reaching to *I* because of the narrow spaces *EFI* and *GHI*. Therefore, being excluded from there, the pressuring matter could not act on the sections *EIG* and *FIH* of the spheres *A* and *C*. Thus, particles *A* and *C*, exposed to the pressing fluid on the rest of their surfaces, must cohere by the measure of the circle or the area of cohesion.

§13

Hence the following rule: *the larger the insensible particles composing a body, the stronger is the coherence, and the smaller they are, the weaker it is.* When the cohereing particles are spheres, then let the semi-diameters of the larger particles [Fig. 1] *AE, CF, AI, CI* = $a$, the radii *EB* and *BF* of the particles of compressing matter = $r$. Then by construction of the figure it is seen that *BI* is perpendicular to *AC*; hence $BI=\sqrt{[(a+r)^2 - a^2]}$. But as *AD, DC, AB, BC* are equal to each other, triangle *ADC* will be = and ~*ABC*; hence *BI* = *DI*; therefore $BD = 2\sqrt{[(a+r)^2 - a^2]}$ = the diameter of the cohesion surface of particles *A* and *C*. Then let $p$ be the circumference of the circle whose diameter = 1; then the area of the cohesion itself will be = $p[(a+r)^2 - a^2]$ [9]. Finally, let the radii of the smaller size particles *A* and *C* = $a - e$ and the semi-diameter of the particles of compressing matter = $r$. And since all can be considered similarly to the above proof, then $BD = 2\sqrt{[(a-e+r)^2 - (a-e)^2]}$ = the diameter of the cohesion surface of the smaller particles, and the area of the cohesion is = $p[(a-e+r)^2 - (a-e)^2]$; thus, the area of cohesion of the larger particles will be to the area of cohesion of the smaller particles = $p[(a+r)^2 - a^2]$ to $p[(a-e+r)^2 - (a-e)^2]$ that is $=(a+r)^2 - a^2$ to $(a-e+r)^2 - (a-e)^2$ or $= r + 2a$ to $r + 2(a-e)$. From this the area of cohesion of the greater particles will be larger than the area of cohesion of the lesser particles; hence the larger the particles, the more strongly they cohere; the smaller they are, the weaker their coherence.

§14

Thus, it is easy to see how many different phenomena in cohesive particles can be explained by this theory that considers the variety of the particles' sizes in the mix. For this reason, let the natural philosopher cease to marvel and doubt that all special qualities of bodies can come from particles which have a spherical shape only, and especially take into consideration the principle of *congruence* of particles presented in the Oration on the origin of light and colors[10]. Beyond this, as an example one can consider the craft by which from round threads, especially if they have different thicknesses, numberless and varied multitudes of textiles and nets are produced in excellent designs by varying their positions.

§15

It has already been shown clearly enough how important is the particles size difference in formation of liquid and solid bodies. Therefore, it remains to consider how the external forces of heat and cold act on different size particles.

§16

From the theory of the rotary calorific motion it follows that the particles of warm bodies spin faster and repel each other with greater force (see Reflection on the causes of heat and cold,

---

[9] Here the original publication [1] had typo "…=$p\sqrt{[(a+r)^2]} - a^2$ …"
[10] See footnote 8 to §9 above

Comm. Nov., T. I, § 23 et 24)[11]; therefore, the cohesion of these particles should become weaker, the greater the warmth or heat is in the body, and it can be enkindled so the body is not only to be converted into a liquid, but further, losing all interosculant cohesion between the particles and their very contact, to be dispersed into a vapor.

§17

Therefore, the bodies with smaller particles require much less calorific to become and stay liquid than for the bodies with larger particles, and it is not surprising that mercury, the fineness and thinness of whose particles have been indicated by many chemical and medical experiments, keeps its liquidity at very low heats which we by our senses call severe cold and strong frost. For, according to the theory of calorific motion, all bodies have heat as long as the particles rotate, although they may seem very cold.

§18

In opposite, bodies of the same nature as mercury, that is, metals, have a stronger coherence between their particles which are larger than those of mercury, and require greater fire to melt them. The greater size of the particles making up metals is clear from that mercury gets into their pores.

§19

There are countless properties and qualities which occur in solid and liquid bodies from different cohesions of their particles, such as different degrees of viscosity, brittleness, softness, friability, ductility, elasticity, and others which require specific and comprehensive considerations and clear understanding of subtle physics; bur instead we, leaving those for the future, will discuss here only how sensible bodies can contract or expand from ebullition to freezing, and contract and expand in reality.

§20

Let's first consider different arrangements of the particles of the spherical shape that can be safely assumed on base of the above arguments. Four spherical contacting particles can be in compact placement in an equilateral rhombic figure, with angles $CB$ equal to 60° and $AD$ 120°, while they should form a cube for the most spatially ample arrangement. Such equilateral figures $ABCD$ (Fig. 2) and $ABCD$ (Fig. 3) have the proportion to each other as $\overline{AB}^3$ to $\frac{1}{2}\sqrt{\overline{AB}^2 + \overline{BC}^2} \times \overline{AB}^2$, that is, as $\overline{AB}^3$ is to $\overline{AB}^2\sqrt{\frac{1}{2}}$ or $= 1:\sqrt{\frac{1}{2}} = 1000$ to $\sqrt{500000}$ because $AC = BC$. For such a rhombic body can be divided into two equal prisms $ADCFBE$ and $ADCFGH$ which have a common square side $ADCF$ [Fig. 4]. And since the angles $ABC$ and $FBD$ are right angles, then the half diagonal *AC* equals the altitude *BK* of the prism *ACDFBE* or half of all the rhombic body. That is, the volume of the cubic body to that of the rhombic body will be almost as 1000 to 707.

---
[11] See footnote 6 to §9 above

§21

Hence it follows: 1) how much simple bodies, those composed of equal particles and without foreign materials in the pores, can be expanded and contracted without losing the cohesion, though it can increase and decrease; 2) that particles of foreign material in the pores, for example, of air, can prevent own particles to attain the ultimate rhombic cohesion and hence cannot always get such a great compression as shown above in §20; however, they still could have sufficient room for contraction and expansion, and one can expect different degrees of contraction and expansion for different amounts of foreign material; 3) as in the cubic arrangement there are twelve contacts between the eight particles, and eighteen in the rhombic one, then it is not surprising that after losing liquidity the particles get into the rhombic arrangement and acquire hard coherence of a solid due to the strength of the six additional contacts.

§22

Different bodies exhibit different contraction in experiments. It can be measured, in all of those which not difficult to freeze and boil, such as in water, in oils, and in solutions of different salts. For the bodies in which freezing has not yet been observed it is impossible to determine the extent of contraction. For example, there was no hope to get it for mercury, for until the last winter it had not been frozen. But now it remains only to resolve by analysis and tests the discrepancies between the observations at what degree below boiling mercury should freeze. What can be concluded from my experiments, is presented below.

§23

On December 26, 1759, when the frost was 208 degrees[12], I put a thermometer in the snow into which I had poured nitric acid; the snow then melted like butter when it is close to its eliquating; the mercury in the thermometer dropped to 330 degrees, then, added new snow and poured mineral acid; the mercury sank to 495 degrees; I added more the acid and saw the mercury at 534 degrees. On taking the thermometer out the mixture for a short time, the mercury reached 552 degrees. Finally, as so-called oil of vitriol[13] was poured to the newly added snow, in an instant the snow got converted into an almost liquid matter and the mercury sank to 1260 degrees[14]. Then, having not doubts that it was already frozen, I quickly struck the sphere with a copper compass and the glass shell was shattered and fell off the mercury ball which remained with a tail reaching into the tube of the thermometer like a pure silver wire which bent freely like a soft metal, and had the thickness of 1/4 of linea[15]. After this, striking the ball of mercury with a butt of ax, I felt that it had a hardness like tin or lead. From the first blow to the fourth it was compressed without a flaw, and from the fifth, sixth and seventh blows it showed cracks [Fig. 5]. *A* shows the ball of mercury with the tail, *B* after the first blow, *C* after the second, *D* after the third and fourth, *E* after the fifth, sixth, and seventh. Then I stopped hammering and started to cut with a knife, and after about 20 minutes it began to pass into an amalgam or into a dough and quickly regained its lost liquidity, that is, melted at such a great frost of 208 degrees. Initially there was no liquid inside, nor were pores noted, and the hardness was greater than in the later experiments. And though, because of haste, I did not see whether there were any cracks in the glass sphere, yet there would be no danger that the mercury would flow out, as the mercury itself already had created a wall when at the first fall in temperature its surface became a solid body and served instead of the vessel to contain that part which had not yet frozen inside.

---

[12] Here and further temperatures are according to the Joseph-Nicolas Delisle scale : the water boiling point was 0 °D, the ice melting point at 150 °D. So, 208 °D is -38.7 ° C and -37.6 ° F.
[13] Sulfuric acid
[14] Formally, 1260 °D is equal to -740 ° C and -1300 ° F. These values do not reflect real temperatures but instead show the degree of fast contraction of the mercury at the freezing point and 1°D corresponds to 1/10000 of the volume change – see details in the commentary.

§24

According to the experiments made in later frosts I observed: 1) that mercury at about 230 degrees begins to thicken somewhat. This can be seen clearly in a narrow bent glass tube, since the mercury itself did not reach equilibrium as fast as at the warm mercury usually did; 2) at about 500 degrees it stops in the tube, but it is for the most part unfrozen in the middle of the bulb or is filled with numerous sizeable pores; 3) in long, narrow glass cylinders or tubes I have observed frozen mercury with obvious gaps *dd* [Fig. 7]; 4) the mercury sometimes drops further when the tube is warmed by hand; 5) we should note here, although it does not exactly relate to this subject, that the electrical force acts through frozen mercury and through glowing iron. The experiment is depicted in Figure 6. *BdeC* is the bent tube containing the mercury immersed in a freezing material, *d* the end of the wire *AB* which is dipped into the mercury and extended to an electrical indicator and heated by candles set on the bottom of glass vessel *H*; *e* is the end of the wire *CF* at the other elbow of the tube, dipping into the mercury and extending to an electrical globe.

§25

From all these experiments and in agreement with the consideration (§20) it follows that: 1) the difference in the freezing point of mercury in the thermometer occurs due to unequal rate of freezing in the thin thermometer tube. For it is natural that small amounts of mercury in there should freeze faster than much larger amounts in the bulb. And thus the frozen mercury in the tube plugs its own way and entirely stanch it up while freezing in the bulb occurs only on the surface and the inner part is entirely liquid and therefore the thermometer does not show a true lower limit of freezing, but stalls at the degree at which the mercury in the tube gets frozen; 2) the ultimate point of freezing for mercury should be at about 1300 degrees; the mercury frozen in the cylindrical tube contained breaks and empty spaces *dd* inside [Fig. 7] while it sank only to 500 degrees, and according to my estimates it would contract further to 1000 degrees if all those empty spaces were filled; moreover, during the first freezing attempt, the mercury had no visible cavities and got frozen solid all through and that's why it sank that low; 3) volumes of bodies in most spacious arrangement and in the most dense one have relative measures as 1000 to 707 (§20), and mercury from its boiling to its freezing according to my observations is compressed by 1647 degrees[16], that is, from 414[17] above zero to 1260 degrees of the thermometer below. Hence the volume decreases about 16/100; then according to the theory there is as much left to contract further, i.e. to almost 3000 degrees. However, such compression may occur quite a while after the freezing of mercury.

---

[15] Russian and English *linea* is equal to 1/10th of an inch

§26

There are still many other liquid bodies which have not become solid in our strongest frosts and do not turn into their type of ice; but these, nevertheless, require detailed studies not lesser than for mercury. According to the duties assigned by Peter the Great to study the riches of nature, our institution will not pass the chance to study them diligently in the future. Divine Providence, advancing the welfare of Russia and the fortunes of our all gracious Autocrat for her everlasting glory will not deprive us of the advances and expansions in the progress of sciences here, as well as of her other institutions set for the Motherland's good and grand solicitudes charged with the general good. Let the zeal and powerful genius of the sons of Russia for the higher sciences get rooted and strengthened under the generous patronage of the great Elizabeth; and let this great day always be the example, modus and encouragement for expression of true advantages and endless gratefulness to the reigning sciences in our country for all future generations.

---

[16] 1647 degrees Delisle correspond to the volume contraction of 1647/10000
[17] -414 degrees Delisle is +708.8 degrees Fahrenheit or +376 degrees Celsius; modern day value of the mercury's boiling point is +674.1 ° F or +356.7 ° C.

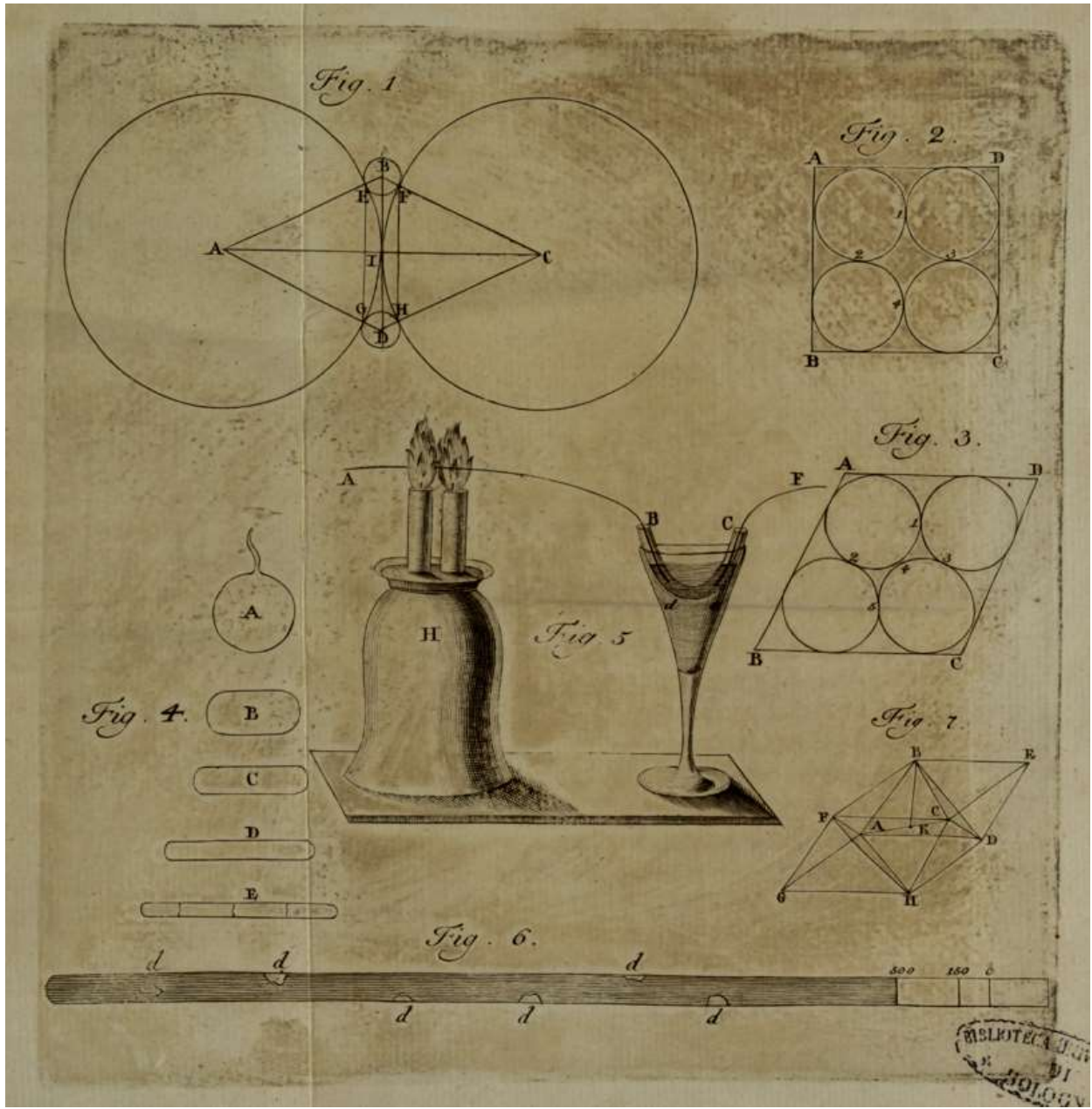

*Commentary* [by V.Shiltsev]:

This English translation of Mikhail Lomonosov's seminal work is made from its Russian original [1]. It continues the series of English translations of Lomonosov's most important scientific works included by himself in the convolute *Lomonosow Opera Academica* sent for distribution among Academies in Europe – the total of nine, see below in the "*List of 26 M.V.Lomonosov's works in natural sciences, history and linguistics, published during his lifetime*". Two other translations of these publications can be found in Refs. [4, 5] and three – "Oration on the Use of Chemistry", "Oration on the Origin of Light" and "Meditations on the Solidity and Fluidity of Bodies" - in the Henry Leicester's book [6]. Unfortunately, the latter, published in [6], pp. 233-246, is found to be unsatisfactory as on numerous instances the meaning of the Russian original was lost or grossly distorted, mostly due to very difficult old Russian language of the Lomonosov's original publication, which is not always easily readable and understood in modern day Russian, making the paper hard to translate to English. In many cases, the Latin version of [1] was needed to come to correct meaning. This text is free of the deficiencies of the H.Leicester's translation that was used only for reference. More on the life and works of the outstanding Russian polymath and one of the giants of the European Enlightenment can be found in books [7, 8] and recent articles [9-14].

The history of this work of Lomonosov, written in August 1760, is as follows. The subjects of extreme cold and of the phase transition from liquid to solid was of significant interest in 18$^{th}$ century. The St.Petersburg Academy of Sciences was on the forefront of that research due to naturally very cold winters, especially in Russia's North and Eastern territories. In mid- to late-1730's, academicians Louis Delisle de la Croyere (1685-1741) and Johann Georg Gmelin (1709-1755) reported from Siberia where they were on decadal scientific Great Kamchatka Expedition, that on a number of occasions they observed congelation of mercury in their thermometers by natural cold – see detailed accounts in [15]. Given uncertainties on the quality of mercury, spread of the freezing temperatures from -25° F to -100° F and absence of confirmations of the effects at some of the reported temperatures under controlled conditions, these observations were usually given little credit [16]. Lomonosov himself carried oou a series of experiments on artificial cold in 1747 (Ref. [2], pp. 411-419), though without a major breakthrough. The events in the St.Petersburg Academy had quickly developed in 1759: first, on November 15 (o.s.) academician August-Nathanael Grischow (1726 – 1760) reported to the Academy's conference yet another observation of the congelation of his thermometer due to natural cold [17]. The freezing of mercury has been undoubtedly established by academician Joseph-Adam Braun (1712-1768) in his experiments on artificial cold in a freezing mixture of snow and acids undertaken on December 14, 1759. His written communication was presented to the Academy's conference on December 20 [18], published in the St.Petersburg newspaper [19] and was later followed by similar confirmatory and exploratory experiments on December 25 and 26 by himself [3], Lomonosov [1], honorary member of the Academy J.G. Model, and academicians Johann Ernst Zeiger (1725-1784) and Franz Ulrich Theodor Aepinus (1724 – 1802). The latter took active part in summarizing the facts and sending brief communications in Europe – the discovery was widely cited, see, e.g., [20-22] – where after explanation of the Braun's experiments, the names of all the epigones and their works were duly cited. Remarkably, Lomonosov has discovered that two Aepinus's communications to Leipzig [23] and Paris [24] were not properly approved by the Academy for publication, suspected that Aepinus made that on purpose to undermine Braun's priority and his own contributions and made a big fuss about that. Unpublished notes and additional paper in support of his position were drawn – see Ref. [2], vol.III, pp.411-427. That ended in nothing but feud between him and Aepinus which further continued into the 1761 transit of Venus affair, see [4, 13].

These experiments led Lomonosov to a series of speculations on the freezing of bodies in which led reviewed and advanced his corpuscular ideas on the nature of cohesion and the structure of liquids and solids. In August 1760 he set these down in a paper, "Meditations on the Solidity and Liquidity of Bodies" [1] which was presented in Russian at the public meeting of the Academy of Sciences on September 6, 1760, right after J.-A.Braun's presentation in Latin [3]. Both papers were soon issued by the Academy as separate publications in Russian and Latin versions quickly followed.

The larger part of the paper [1], §§1-22 is dedicated to theoretical views: Lomonosov repeats the arguments against the existence of an innate attractive force in bodies, shows mathematically that the spherical form was the only one in which material particles could exist, and calculates the surface of cohesion in a mechanical model for the spheres contacting at one point only while being surrounded by a small particles of some *pressing matter*. The resulting pressure to cohere is found proportionally greater for larger spheres, and, thus, bodies with larger corpuscles would show greater cohesion, and, correspondingly, higher melting points. Mercury, known to penetrate into other materials and, therefore, thought to be consisting of smaller corpuscles should then have very low freezing temperature. Lomonosov's simple mathematical model of contacting spheres predicts that the density of the most tightly packed corpuscles after freezing and that of the barely contacting particles at boiling should differ by factor $1/\sqrt{2} = 0.707$.

The last paragraphs §§23-26 present details of the Lomonosov's experiments on freezing of mercury which he undertook on very cold days of December 25 and 26, 1759, some of them quite innovative but using Braun's recipe for the artificial cold (i.e., additional cooling in the mixture of snow and various acids). Of importance are the experimental proofs of the metallic nature of mercury, such as its elasticity and electric conductivity. The most notable controversy, discussed for a quarter of century until similar but better arranged experiments by Hutchins, Cavendish, Black and McNab in 1783-1788 [25-27], was the actual freezing temperature, the modern value of which is -38.9°C or -38.0°F. Lomonosov and Braun used the thermometers originally proposed and developed since 1732 by their fellow academician Joseph-Nicolas Delisle (1688-1768). The Delisle thermometers used mercury as the material with large coefficient of thermal expansion and were calibrated such that 1°D corresponded to 1/10,000 of the mercury volume change. Correspondence to other temperature scales is linear within the working ranges and, e.g., the temperature of boiling water is set 0°D and the freezing temperature of water is 150°D (see more discussion on usefulness of such inverted scale in [16]). Thus, the freezing mercury should be observed at 208°D.

In reality both researchers had winessed indications of the frozen mercury over the range of the temperature readings starting from 208°D to 1260°D (Lomonosov) and even 1500°D (Braun). They had somewhat different attitudes to their highest (coldest) values – while both tried to find and exclude experimental deficiencies (purity of mercury, attention to accounting of mercury if it was lost through the cracks in glass, etc), Braun had no understanding of the high temperatures and left generally suspicious, while Lomonosov was puzzled that the contraction was not sufficiently large to match the prediction of his theory (0.707 or almost 3000°D to freezing below the mercury boiling point of -385°D or +356.7°C). Still, both researchers started to observe the signs of congelation of mercury close to the modern value of 208°D.

As pointed by V.Bilyk [28], explanation of the observed phenomenon was in a quick contraction of mercury at the freezing accounting for 3.7% increase in density during the phase transition. It is remarkable that Braun and Lomonosov obviously did not know about that effect, while the phenomenon was and is one of the basic in the casting – for example, the corresponding solidification shrinkages are 2.3% for lead, 3% for steel, 4.9% for copper, etc. Even though the sign is opposite for water, some 8% expansion occurs at freezing to ice, the significance of the effect should had been notable. We need to note here, though, that the 3.7% contraction in mercury is equivalent to 370°D change and thus would result in the spread of the reported temperatures from 208°D (true point of freezing of the thin outer layer of the mercury in the thermometer bulb) to 208+370 = 578°D. Indeed, both Broun and Lomonosov noted that some intermediate values of temperature as well as other strange phenomena (dependence on shaking or effects of quick warming) were probably due to incomplete freezing of the bulk of mercury. Still, even now it is not fully clear for us how the most extreme readings above 578°D were possible, as not many other clues were left in the reports except to blame possible uncontrolled leakage of mercury through minor cracks in the glass of the thermometers. Experimental replication of Braun's and Lomonosov's tests – like the one on the discovery of Venus's atmosphere [11] - would greatly help to resolve this puzzle.

Lomonosov's simplistic theoretical prediction of the ratio of densities of the most packed and the least packed arrangements of corpuscles to be 0.707 also did not stand out. E.g., the density of mercury at absolute zero is now calculated to be 16.39, while the density at the boiling point is about 13.53; thus, the ratio in fact is 13.53/16.39=0.825 – i.e., much less than 0.707.

Let us note one more subtlety of the experiment shown in Lomonosov's Fig.6. Part of the electric circuit between solid mercury in the curved glass tube and an electrometer is an iron wire heated by two candles. Lomonosov makes an unambiguous conclusion - "The electric force acts through the frozen mercury and through the red-hot iron."

Fig.7 presumably shows glass tube filled with frozen mercury and its scale reads 500°D, that should be equal to 5% linear contraction of the mercury compared to that at the water boiling point of 0°D. Actual measurement of the image indicates 10%, or twice the contraction. The discrepancy can be explained either by engraver's mistake or, most probably, by the fact that the glass tube was curved (as shown on Fig.6 above) and laid on a table, so only *one half* of the tube is visible.

Finally, in §8 of the "Meditations on Solidity and Fluidity of Bodies" Lomonosov, for the first time in a published paper, formulates the conservation of matter and motion as "the universal law" of nature. Twelve years earlier, on July 5, 1748, this law was set out in a letter to Leonhardt Euler (see PSS [2], vol.II, pp. 182-185) - "…all the changes that occur in the species occur in such a way that if something is added to something, it is taken away from something else. So, how much matter is added to any body, as much is lost in the other". Later it was repeated in 1758 in an unread speech on the relation of the amount of matter and weight (PSS [2], vol. III, pp. 349-371). Soviet scholars somewhat hyperbolically often credited Lomonosov with the discovery of the general conservation laws. That surely is not correct. But despite skepticism and criticism of other scholars - see , eg. [29] – it is still hard to deny that at least Lomonosov was first - in 1756, seventeen years prior to analogous results by Lavoisier - to experimentally demonstrate the law of conservation of the weight of a substance in chemical reactions to a high degree of accuracy by showing that lead plates in a sealed vessel without access to air do not change their weight after heating and calcination (see his internal report to the President of the St.Petersburg Academy in PSS [2], vol. X, pp. 392).

I would like to thank Prof. Robert Crease of SUNY, my long-term collaborator and co-author of several scholar papers on Mikhail Lomonosov, for the encouragement to translate Lomonosov's major works to English.

*великия государыни императрицы Елисаветы Петровны самодержицы всероссийския, в публичном собрании императорской Академии Наук сентября 6 дня 1760 года читанное господином коллежским советником и профессором и королевской шведской Академии Наук членом Михаилом Ломоносовым*" (СПб., при имп. Академии Наук, 1760, 21 стр); in Latin – in " *Meditationes de solido et fluido solemnibus sacris augustissimi nominis serenissimae potentissimae magnae dominae imperatricis Elisabethae, Petri Magni filiae, autocratoris omnium Rossiarum, praelectae in Academiae Scientiarum Petropolitanae conventu publico die VI Sept. A. MDCCLX. Auctore Michaele Lomonosow, consiliario academico, chymiae professore nee non regiae Academiae Scientiarum Sueciae membro* (Petropoli, typ. Academiae, 1760, 18 pages); both texts are available in *Lomonosov Complete Works* [2], v.3, pp.377-409.

**Appendix: List of M.V.Lomonosov's works in natural sciences, history and linguistics, published during his lifetime**

**In bold** – publication, out of total of 9, included by M.V.Lomonosov in his convolute *Lomonosow Opera Academica* sent for distribution abroad.

By asterisk * - publications noted by M.V.Lomonosov in his 1764 internal report *Review of the most important discoveries with which Mikhailo Lomonosov has tried to enrich the natural sciences*. (PSS [2], v. 10, pp.404-411)

| *Title* | *Year Written* | *Year Published* | *PSS Vol.* | *pp.* |
|---|---|---|---|---|
| De motu aeris in fodinis observato. Auct. M. Lomonosow. О вольном движении воздуха, в рудниках примеченном. [Русский перевод Ломоносова] | 1742-44 | 1750 | I | 315-331 |
| * Dissertatio de actione menstruorum chymicorum in genere. Auctore M. Lomonosow [О действии химических растворителей вообще, Михаила Ломоносова] | 1743 | 1750 | I | 337-383 |
| * De tincturis metallorum. Auctore Michaele Lomonosow. О металлическом блеске. Михайло Ломоносов. | 1745 | 1751 | I | 390-417 |
| Волфианская экспериментальная физика, с немецкого подлинника на латинском языке сокращенная, с которого на российский язык перевел Михайло Ломоносов, императорской Академии Наук член и химии профессор. | 1745 | 1746 | I | 419-536 |
| * Meditationes de caloris et frigoris causa. Auctore Michaele Lomonosow. [Размышления о причине теплоты и холода Михаила Ломоносова] | 1749 | 1750 | II | 7-55 |
| * Tentamen theoriae de vi aeris elastica. Auctore Michaele Lomonosow. [Опыт теории упругости воздуха Михаила Ломоносова] | 1748 | 1750 | II | 105-139 |
| Supplementum ad meditationes de vi aeris elastica. Auctore Michaele Lomonosow. [Прибавление к размышлениям об упругости воздуха Михаила Ломоносова] | 1749 | 1750 | II | 145-163 |
| Anemometrum summam celeritatem cujusvis venti et simul variationes directionum illius indicans. Auctore Michaele Lomonosow. [Анемометр, показывающий наибольшую быстроту любого ветра и одновременно изменения в его направлении, Михаила Ломоносова] | 1748 | 1751 | II | 205-217 |

**bei der Kaiserlichen Akademie der Wissenschaften: Aus dem Russischen übersetzt. St. Petersbourg**

| | | | | |
|---|---|---|---|---|
| Catalogus minerarum [Минеральный каталог/Каталог камней и окаменелостей Минерального кабинета Кунсткамеры Академии Наук] | 1741 | 1745 | V | 7-241 |
| * **Слово о рождении металлов от трясения земли. Oratio de generatione metallorum a terrae motu** | 1757 | 1757 | V | 295-347 |
| Первые основания металлургии или рудных дел | 1742 | 1763 | V | 397-631 |
| [Древняя Российская история от начала российского народа до кончины великого князя Ярослава Первого или до 1054 года, сочиненная Михаилом Ломоносовым, статским советником, профессором химии и членом Санкпетербургской императорской и королевской Шведской Академий Наук] | 1754 | 1766 | VI | 163-286 |
| Краткий Российский летописец с родословием. Сочинение Михаила Ломоносова | 1759 | 1760 | VI | 287-358 |
| Краткое руководство к красноречию. Книга первая, в которой содержится риторика, показывающая общие правила обоего красноречия, то есть оратории и поэзии, сочиненная в пользу любящих словесные науки | 1744 | 1748 | VII | 89-378 |
| Российская грамматика | 1754-55 | 1757 | VII | 389-578 |
| Предисловие о пользе книг церковных в российском языке | 1758 | 1758 | VII | 585-592 |